\documentstyle[epsf,psfig]{mn}

\title{Photopolarimetric observations of the new eclipsing
polar CTCV J1928-5001. }

\author[Stephen\,B. Potter et al.] {Stephen\,B. Potter$^{1}$, Thomas
Augusteijn$^{2}$ \& Claus Tappert$^{3}$\\
$^{1}$South African Astronomical Observatory, PO Box 9,
Observatory 7935, Cape Town, South Africa\\ 
$^{2}$Nordic Optical Telescope, Apartado 474, 38700 Santa Cruz de La
Palma, Canary Islands, Spain \\
$^{3}$Departamento de Astronom\'{\i}a y Astrof\'{\i}sica, Pontificia
Universidad Cat\'olica, Casilla 306, Santiago 22, Chile\\
}

\date{}

\begin{document}

\maketitle

\begin{abstract}

We report photopolarimetric observations of a new eclipsing polar (AM
Herculis system) discovered in the Cal\'an-Tololo survey. The
photometry and polarimetry are modulated on a period of $\sim$ 101
minutes. Circular polarization variations are seen from $\sim -8$ to
$+12$ per cent and from $\sim 0$ to $5$ per cent in the red and blue
parts of the optical spectrum, respectively.  Two linearly polarized
pulses are detected at orbital phases coinciding with the reversals in
the circular polarization.  This is consistent with a magnetic field
strength of $\sim$ 20 MG for the white dwarf primary, where accretion
takes place at two regions. Both accretion regions are self-occulted
by the white dwarf during parts of the orbit.  We estimate some of the
system's parameters from its eclipses, which we further refine by
modelling the polarimetric observations.

\end{abstract}

 \begin{keywords}
     accretion, accretion discs -- methods: analytical -- techniques:
     polarimetric -- binaries: close -- novae, cataclysmic variables --
     X--rays: stars. 
 \end{keywords}

\section{Introduction}

Polars (AM Herculis systems) are short period binaries that contain a
white dwarf (known as the primary star) and a red dwarf (the secondary
star) that overflows its Roche lobe. The magnetic field of the white
dwarf is sufficiently strong to lock the system into synchronous
rotation and to prevent an accretion disc from forming. Instead, the
overflowing material from the secondary star initially continues on a
ballistic trajectory until, at some distance from the white dwarf, the
magnetic pressure overwhelms the ram pressure of the ballistic
stream. From this point onwards the accretion flow is confined to
follow the magnetic field lines of the white dwarf, see e.g. Cropper
(1990) or Warner (1995) for a full review of magnetic cataclysmic
variables (mCVs).

In high inclination systems, the secondary star occults the white
dwarf, the accretion region and the magnetic and ballistic accretion
streams, which produces structured photometric eclipse profiles. Thus
eclipses offer unique opportunities to investigate these spatial
structures and to constrain the binary parameters (e.g. HU Aqr: Hakala
et al. 1993 and Bridge et al. 2002, UZ For: Imamura, Steiman-Cameron
\& Thomas 1998 and V1309: Katajainen et al. 2003). In addition, the
polarimetric variations can give further insights into the accretion
region. With appropriate modelling, one can estimate the shape, size
and location of the accretion region on the surface of the white dwarf
(e.g. Bailey et al. 1995 and Potter, Hakala \& Cropper 1998), which
aids the determination of the magnetic field structure of the white
dwarf and the extent of the ballistic and magnetic streams. Knowing
the location of the accretion region also helps to interpret
spectroscopic observations: the cyclotron and bremsstrahlung radiation
emitted from the accretion region is reprocessed by the other binary
components and is seen as, for example, modulated emission line
variations (see e.g. Schwope et al. 2000, Romero--Colmenero et
al. 2003).

CTCV~J1928-5001 was identified spectroscopically and photometrically
in the Cal\'an-Tololo survey (see Tappert, Augusteijn \& Maza 2004) as
an AM Her system candidate. In this paper, our follow-up high time-resolved
photometry has confirmed its CV characteristics with an orbital
modulation of $\sim$ 101 minutes interspersed by deep, short eclipses.
The optical light-curve is characterised by a double-humped bright
phase and a faint phase. The eclipse ingresses and egresses are
particularly sharp, typically 2-3 seconds in duration, which have
enabled us to calculate an accurate eclipse ephemeris. The eclipse
itself is relatively short in duration, lasting $\sim $ 177
seconds. We discuss the shape of the eclipse profile and the
constraints that it places on the binary parameters, such as its
inclination, mass-ratio and white dwarf mass.

In addition, we use Stokes imaging (see Potter, Hakala \& Cropper 1998
and Potter et al. 2004) to model the polarimetric observations and we
show that accretion takes place at two regions on the primary.  We
place further constraints on the binary parameters by combining the
analysis of the eclipses and the polarimetric modelling with simple
single particle trajectory modelling.

\section{Observations}

\subsection{Photometry}

\begin{figure}
\epsfxsize=8.5cm
\epsffile{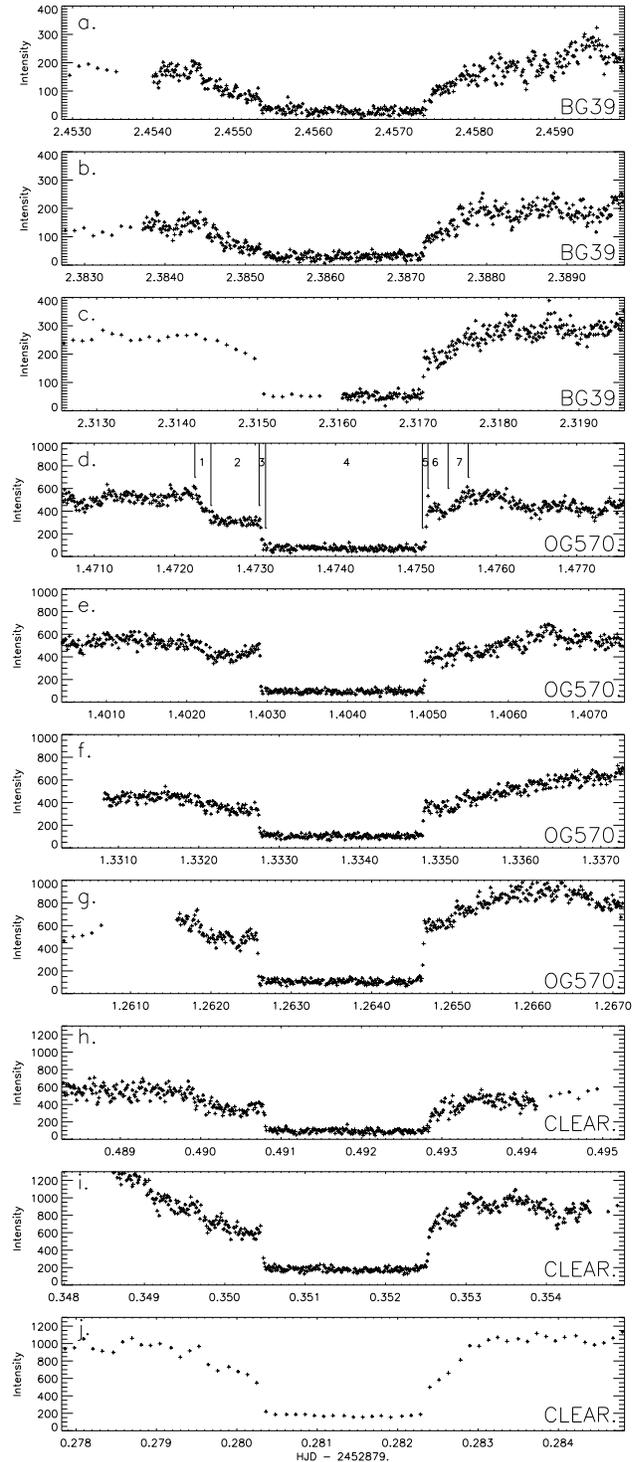} 
\caption{The eclipses centered on the orbital period. a,b and c are
taken through a BG39 filter, d,e,f and g are taken through an OG570
filter and h,i and j are unfiltered.}
\label{alleclipses}
\end{figure}

The 2003 photometric observations were obtained as part of the
polarimetric observations. Over the course of three nights, unfiltered
(defined by an RCA31034A GaAs photomultiplier response 3500 -- 9000
\AA), OG570 filtered and BG39 filtered light curves were
acquired. Conditions were photometric during the whole observing
campaign. The data were sky subtracted and extinction corrected, but
not flux calibrated. From the observed count rate, however, we
estimate that the system was caught in a brighter state (V magnitude
$\sim$ 16-17) than the 1996 observations (Tappert, Augusteijn \& Maza
2004). Figs.~\ref{alleclipses}, \ref{unfilt}, \ref{OG570} \&
\ref{BG39} show the eclipse profiles and the folded light curves
respectively.

\begin{table*}
\begin{center}
\caption{Log of observations. `1.9m SAAO' is the 1.9m telescope of the
South African Astronomical Observatory. `UCTPol' is the University of
Cape Town photopolarimeter. `Eclipses' indicates the number of
eclipses each observation covered. `10 + 1s' means that we obtained 10
and 1 second time resolution out and during eclipse,
respectively. {\label{tab:observations}}}
\vspace{0.2cm}
\centerline{ 
\begin{tabular}{|l|c|c|c|c|l|} \hline
date & telescope & instrument &filter & time resolution & eclipses \\
\\ 
27 Aug 2003   & 1.9m SAAO & UCTPol & Unfiltered  & 10 + 1 s &  3\\
28 Aug 2003   & 1.9m SAAO & UCTPol & OG570  & 10 + 1 s & 4\\
29 Aug 2003   & 1.9m SAAO & UCTPol & BG39  & 10 + 1 s &  3\\
\\
\hline\hline 
\end{tabular}
}
\end{center}
\end{table*}

\begin{table*}
\begin{center}
\caption{Table of eclipse timings. Ingress, egress and mid-eclipses are
HJD-2450000. The 1996 eclipses were recalculated from the observations
of Tappert, Augusteijn \& Maza 2004 (see text).  {\label{tab:eclipses}}}
\vspace{0.2cm}
\centerline{ 
\begin{tabular}{|l|r|r|r|r|} \hline
Date       & Ingress        & Egress        & Mid-eclipse         & Cycle\\
\\ 
8 Jul 1996 & 273.7286(32)   & 273.7382(32)  &  273.7334(55)   & -37136\\
8 Jul 1996 & 273.8016(13)   & 273.8056(13)  &  273.8036(23)   & -37135\\
9 Jul 1996 & 274.7135(11)   & 274.7179(11)  &  274.7157(19)   & -37122\\
9 Jul 1996 & 274.7846(11)   & 274.7878(11)  &  274.7862(19)   & -37121\\
9 Jul 1996 & 274.8538(11)   & 274.8592(11)  &  274.8565(19)   & -37120\\
30 Jul 1996& 295.62312(69)  & 295.6259(7)   &  295.6245(12)   & -36824\\
30 Jul 1996& 295.69125(69)  & 295.6973(7)   &  295.6947(12)   & -36823\\
27 Aug 2003& 2879.28030(12) & 2879.28234(12)&  2879.28132(16) & 0\\
27 Aug 2003& 2879.350472(12)& 2879.352527(12)& 2879.351500(16)& 1\\
27 Aug 2003& 2879.490793(12)& 2879.492859(12)& 2879.491826(16)& 3\\
28 Aug 2003& 2880.262589(12)& 2880.264638(12)& 2880.263614(16)& 14\\
28 Aug 2003& 2880.332752(12)& 2880.334801(12)& 2880.333777(16)& 15\\
28 Aug 2003& 2880.402916(12)& 2880.404953(12)& 2880.403934(16)& 16\\
28 Aug 2003& 2880.473073(12)& 2880.475116(12)& 2880.474095(16)& 17\\
29 Aug 2003& 2881.31496(12) & 2881.317063(12)& 2881.31601(12) & 29\\
29 Aug 2003& 2881.385212(12)& 2881.387214(12)& 2881.386213(16)& 30\\
29 Aug 2003& 2881.455323(12)& 2881.457407(12)& 2881.456366(16)& 31\\
\\
\hline\hline 
\end{tabular}
}
\end{center}
\end{table*}

\begin{table*}
\begin{center}
\caption{Table of system parameters. The mass of the secondary is
  assumed to be 0.1M\sun \ and the eclipse length to be $\Delta \phi =
  0.029$. $\beta$ is the angle between the magnetic axis and the spin
  axis of the white dwarf. The range in the white dwarf radius is estimated
  from In-Saeng Suh \& Mathews (2000) {\label{tab:sysparms}}}
\vspace{0.2cm}
\centerline{ 
\begin{tabular}{|c|c|c|c|c|} \hline
Inclination (degrees) & $\beta$ (degrees) & mass ratio (M2/M1)& White dwarf mass/M\sun & White dwarf
radius $\times$ 100/R\sun \\
\\ 
70 & 29 & 0.97 & 0.10  & 1.7 - 2.3 \\
74 & 23 & 0.42 & 0.24  & 1.3 - 1.7 \\
78 & 18 & 0.20 & 0.50  & 1.0 - 1.3 \\
82 & 12 & 0.07 & 1.43  &    $<$ 1.0 \\
\hline\hline 
\end{tabular}
}
\end{center}
\end{table*}

\subsection{Polarimetry}

\begin{figure}
\epsfxsize=8.5cm
\epsffile{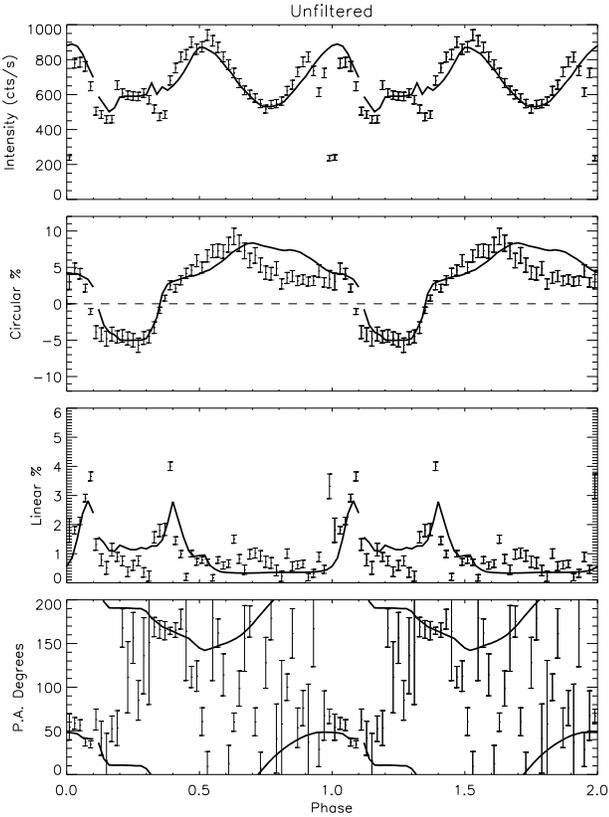} 
\caption{The polarimetric data (white light) folded on the eclipse
ephemeris derived in section 3.2. Solid curves are the model fit (see
section 3.6) }
\label{unfilt}
\end{figure}

\begin{figure}
\epsfxsize=8.5cm
\epsffile{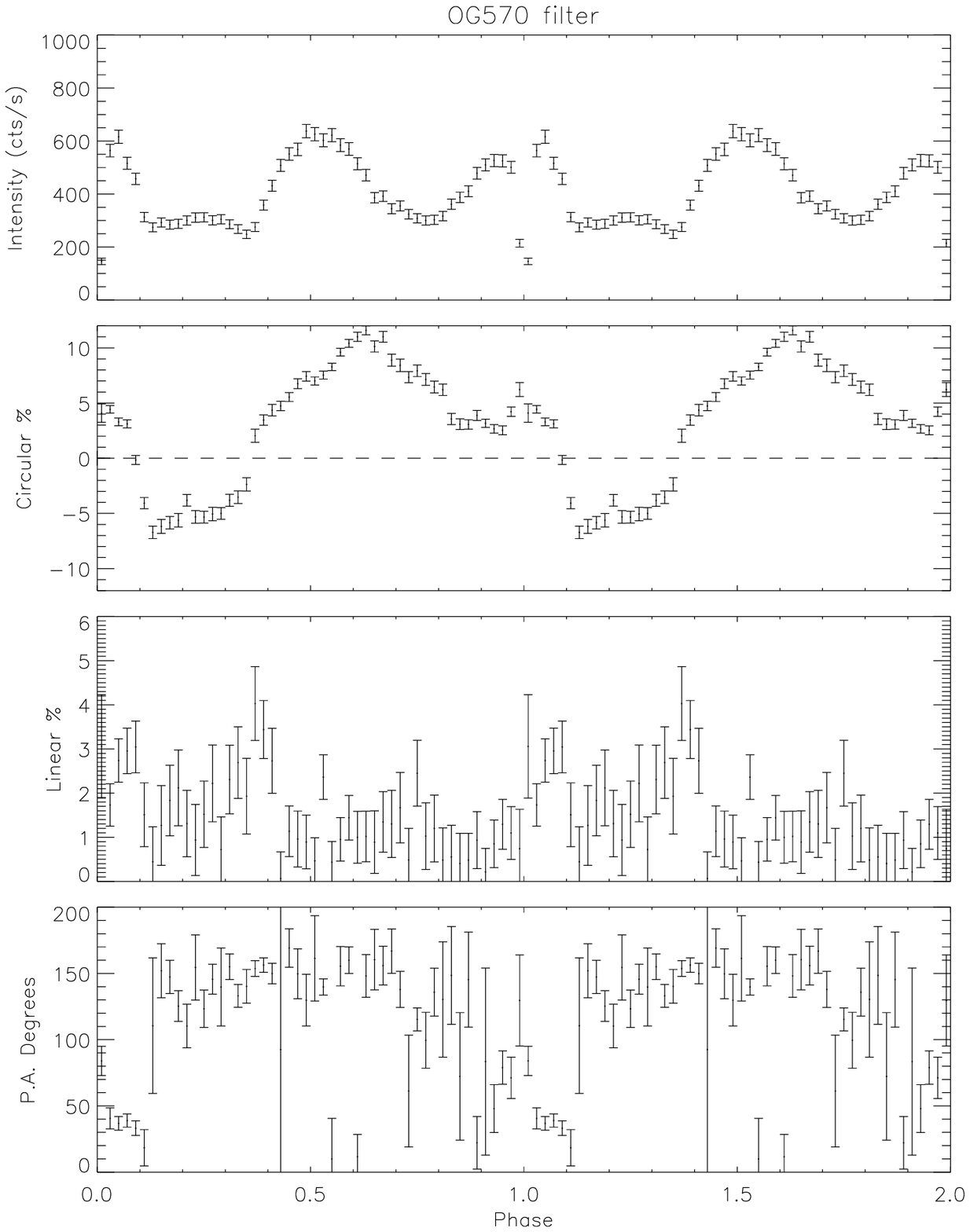} 
\caption{The polarimetric data (OG570 filter) folded on the
eclipse ephemeris derived in section 3.2. }
\label{OG570}
\end{figure}

\begin{figure}
\epsfxsize=8.5cm
\epsffile{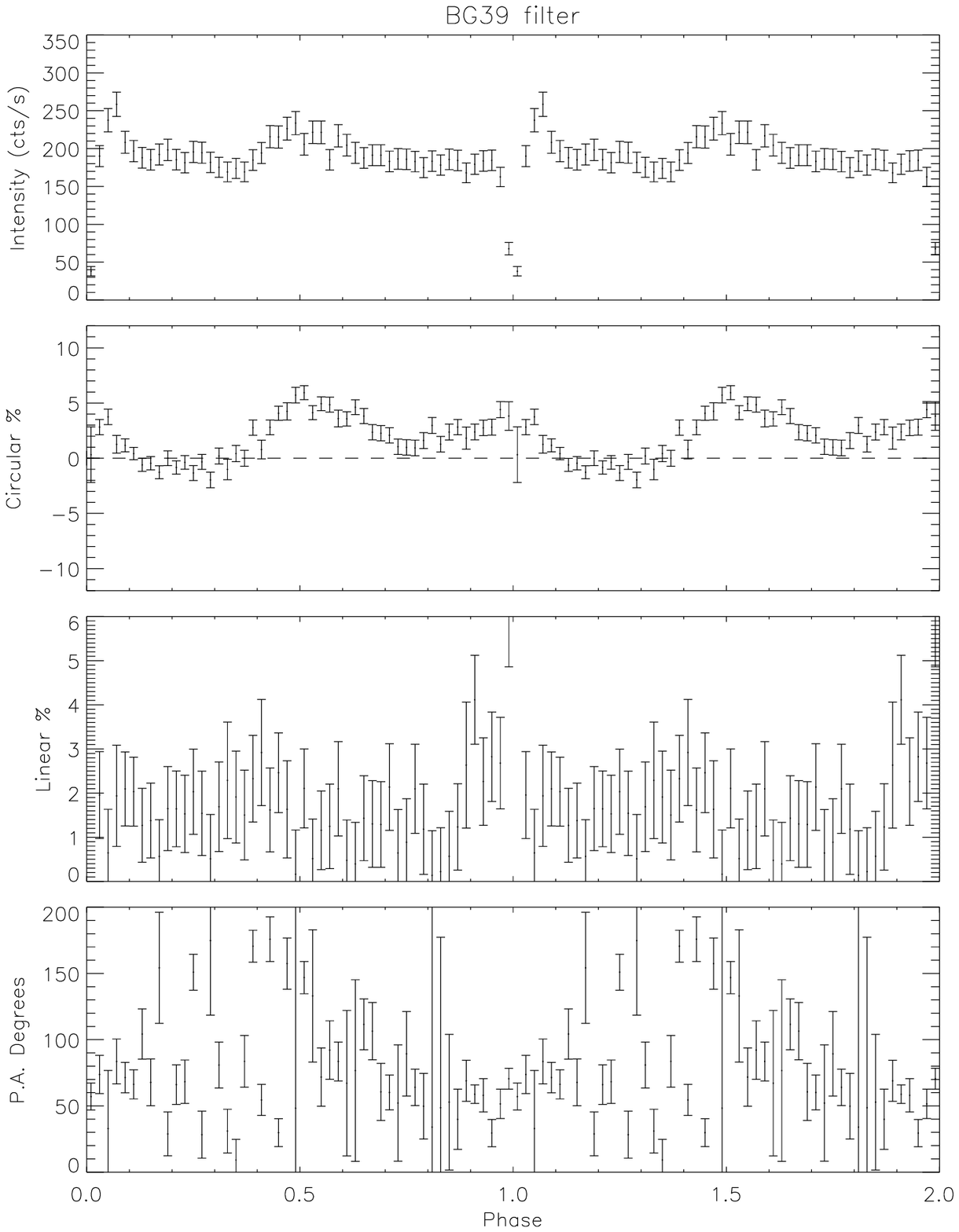} 
\caption{The polarimetric data (BG39 filter) folded on the
eclipse ephemeris derived in section 3.2. }
\label{BG39}
\end{figure}

CTCV~J1928-5001 was observed on three nights during August 2003 (see
table ~\ref{tab:observations}) using the South African Astronomical
Observatory (SAAO) 1.9-m telescope and the UCT polarimeter (UCTPol;
Cropper 1985). The UCTPol was operated in Stokes mode,
i.e. simultaneous linear and circular polarimetry, and
photometry. Broad filtered (BG39 and OG570 covering 3500 -- 5500 \AA \
and 5700 -- 9000 \AA \ respectively) and white light (defined by an
RCA31034A GaAs photomultiplier response 3500 -- 9000 \AA),
observations were taken (see Figs. \ref{unfilt}, \ref{OG570} \&
\ref{BG39}).

Polarised standard stars (Hsu \& Breger 1982) were observed during the
nights to set the position angle offsets.  Non-polarized standard
stars and calibration polaroids were observed to set the efficiency
factors. Background sky polarisation measurements were also taken at
frequent intervals during the observations. Polynomial fits to the sky
measurements were subtracted from the object measurements. The data
were reduced as described in Cropper (1997).

\section{Analysis}

\subsection{The eclipses}

Fig.~\ref{alleclipses} shows the ten eclipses observed in 2003. Upon
inspection of the 2003 eclipses, we identify seven features which vary
from one eclipse to another and may be wavelength-dependent. These
features are indicated in Fig.~\ref{alleclipses}d, where they appear
most pronounced. They are also listed below in order of increasing
time, and we interpret them using geometric arguments based on the
standard picture of an mCV.

1. There is first a slow decrease in brightness of length $\sim 15-20 s$. We
   attribute this to the gradual ingress of the stream beginning at the inner
   Lagrangian point.

2. Next, there is a plateau of almost constant brightness for $\sim
   40-45s$. With the stream now mostly eclipsed, the plateau can be
   interpreted as emission from the photosphere of the white dwarf and
   from the accretion region. This is further supported by the
   continued presence of circular polarization during this period (not
   shown). In some cases (e.g. Fig.~\ref{alleclipses} a,b and i) the
   plateau is not so obvious; this may indicate that the extent and/or
   brightness of the stream changes over time.

3. There is next a rapid drop in brightness to $\sim$20\% of maximum
   intensity which takes only $\sim 3$ seconds. The phase of the rapid
   drop appears to be very stable between the observed eclipses (see
   Fig.~\ref{alleclipses}) and the magnitude of the dip appears to be
   wavelength-dependent. Therefore, we attribute this feature to the
   ingress of the white dwarf and the accretion region.  The data is
   of insufficient time resolution and/or signal-to-noise to resolve the
   eclipse of the photosphere of the white dwarf and the accretion
   region separately. (see e.g. the eclipse of UZ For: Imamura,
   Steiman-Cameron \& Thomas 1998).

   In general, the trajectory of the stream is expected to result in
   the stream lagging the white dwarf and, consequently, eclipsing
   later than the white dwarf and the accretion region. Therefore, we
   would expect to see the phase of slower decline in brighness to
   follow the rapid drop (e.g. HU Aqr: Hakala et al. 1993 and Bridge
   et al. 2002, UZ For: Imamura, Steiman-Cameron \& Thomas 1998 and
   V1309: Katajainen et al. 2003). Since this phase is before the
   rapid drop, we reason that, in this system, the brightest part of
   the stream is mostly eclipsed by now, and there is little or
   insignificant emission from the latter parts of the accretion
   stream.

4. Next, the light curve takes on a flat minimum, which can be
   associated with the complete eclipse of the white dwarf and the
   accretion stream.  This feature lasts just under three minutes.

5. The total eclipse is terminated by a rapid increase in intensity
   ($\sim 500$\%) in $\sim 3-4$ seconds. We attibute this to the
   egress of the white dwarf and the accretion region. Again, as with
   the ingresses, the egresses appear to be stable in phase between
   eclipses.  These data are also of insufficient time resolution and/or
   signal-to-noise to resolve the photosphere of the white dwarf and the
   accretion region separately.

6. Next follows a short plateau of constant brightness lasting $\sim
   15-20s$. This is most likely due to emission from the photosphere
   of the white dwarf and the accretion region. This is again further
   supported by the presence of circular polarization during this
   period (not shown). As with the plateau before ingress (note 2
   above), the length of this post-eclipse plateau is not so obvious
   in some cases. This again may indicate that the extent and/or
   brightness of the stream changes over time.

7. Finally, the intensity slowly increases to the pre-eclipse level, as
   the brighter parts of the accretion stream gradually come into
   view.

\subsection{An eclipse ephemeris}

We re-analyzed the observations of Tappert, Augusteijn \& Maza (2004)
and discovered that, due to a software error, part of their dataset
had not been heliocentric corrected, leading to an erroneous
ephemeris. We therefore re-calculated the times of the ingresses and
egresses of the 1996 observations. Since the 1996 eclipses contained
unresolved ingresses and egresses, the uncertainty in these
measurements was taken to be the time resolution in the
data. Mid-eclipse times were then calculated as the average of the the
ingress and egress for a given eclipse. The times of mid-eclipses with
their uncertainties are shown in (table~\ref{tab:eclipses}).

We used the mid-point between the extremely sharp ($\sim 3$ seconds)
ingresses and egresses of the 2003 eclipses to calculate the times of
mid-eclipse (see table~\ref{tab:eclipses}). A least squares polynomial
fit (order 1) gives an orbital period of $0.0701623$d with an error of
$5.7 \times 10^{-7}$d. The six year gap between the 1996 and the 2003
observations correspond to $\sim 37000$ orbital cycles, giving an
accumulated error of $\sim 30$ minutes. This is sufficiently small to
assign unambiguously cycle counts to the 1996 eclipses, as well as to
those observed in 2003. A least squares polynomial fit (order 1) to
the times of mid-eclipse of both datasets together yields the
following eclipse ephemeris:

$$T(HJD) = 2452879.2813386(58) + 0.070162312(9)E$$

From the 2003 eclipses alone, we calculate the average of the total
eclipse time to be $177\pm 2.3$ seconds, which corresponds to $\Delta
\phi = 0.0292 \pm 0.0004$ of the orbit. The error is estimated from
the standard deviation.

\subsection{The photometry}

Phase-folded photometric light-curves are presented in
Figs. \ref{unfilt}, \ref{OG570} \& \ref{BG39}. 


As can be seen from these figures, there is a bright phase occurring
just before the eclipse. In contrast, the bright phase is more
centered on the eclipse during the low state 1996 observations (see
Tappert, Augusteijn \& Maza 2004). A similar effect has been seen in
other eclipsing polars, e.g. HU Aqr (Schwope et al. 2001), and can be
explained as the relocation of an accretion region on the surface of
the white dwarf between high and low states. During the low state, the
ballistic stream is thought to attach to magnetic field lines soon
after leaving the inner Lagrangian point. Consequently, the accretion
region forms relatively close to the line of centers of the two stars,
causing the bright phase to be centered approximately on the
eclipse. During the high state, however, the ballistic stream has more
ram pressure, so it is able to penetrate further around the white
dwarf before attaching to magnetic field lines. As a result, the
accretion region is located further in longitude from the line of
centers of the two stars, and the center of the bright phase then
occurs sometime before the eclipse.

The double-humped nature of the bright phase is discussed in the next
section.

\subsection{The polarimetry}

Figs.~\ref{unfilt},\ref{OG570} \& \ref{BG39} also show the unfiltered,
OG570 \& BG39 polarimetric observations, respectively, folded and
binned on the orbital ephemeris derived in section 3.2. As can be seen
from these figures, the polarimetry shows both positive and negative
circular polarization. This is typical of accretion onto two regions
on the surface of the white dwarf which are located close to magnetic poles of
opposite magnetic polarity (e.g. Buckley \& Shafter 1995).

The bright phase in the photometry and polarimetry show the typical
morphological features of cyclotron emission from a magnetically
confined accretion shock on the surface of the white dwarf, namely a
double-humped variation due to the angular dependence of the beaming
of the cyclotron radiation, and a wavelength dependence. Most of the
polarized flux is seen through the red (OG570) filter, with the
circular polarization rising to 12 percent. Polarized flux rises to a
maximum of 5 percent through the blue (BG39) filter. This is typical
of polars with magnetic field strengths of about $\sim 20MG$ (e.g. ST
LMi: Peacock et al. 1992). This magnetic field strength is consistent
with the Zeeman splitting of the white dwarf absorption lines reported
in Tappert, Augusteijn \& Maza 2004.

The bright phase lasts $\sim 0.75$ of an orbit, indicating that the
main accretion region is located in the hemisphere closest to Earth
(say, the upper hemisphere). As a rough starting point for Stokes
Imaging, we have used this value to estimate the magnetic dipole
offset angle ($\beta$) as a function of inclination using equation 1
of Visvanathan \& Wickramasinghe (1981).  This equation assumes that
the accretion region is not extended and that it is located at the
magnetic pole. We argue that this is a good starting point, however,
because one would expect to see a more pronounced asymmetry in the
circularly polarised light curves and in the linearly polarized pulses
if the accretion region was significantly extended or offset from the
magnetic pole (see the cyclotron emission models of Potter 1998 and
Ferrario et al. 1990, and see Bailey et al. 1995 and Ramsay et
al. 1996 for examples of systems where the accretion region is thought
to have a large offset from the magnetic pole).  The range of
inclinations is estimated using the observed eclipses and is discussed
in section 3.5. The results are shown in table~\ref{tab:sysparms}.

During the faint phase, negative circular polarization is seen in the
unfiltered and red (OG570) filter only
(Figs. ~\ref{unfilt},\ref{OG570} and, conversely, Fig.~\ref{BG39}),
rising to a maximum of $\sim$ (minus) 7 percent. The faint phase in
the photometry and polarimetry lasts $\sim 0.25$ of an orbit,
indicating that the second accretion region is located in the
hemisphere of the white dwarf farthest from Earth (the lower
hemisphere). Although it is possible that the lower accretion region
is in view for a longer period than that indicated by the negative
circular polarization, there is no evidence of this in our data
(e.g. linear pulses at phases different from those described below).

There are two linearly polarized pulses centered on phases $\sim 0.4$
and $\sim 0.05$, most evident in the unfiltered and OG570 observations
(Figs.~\ref{unfilt} \& \ref{OG570}). These coincide with the reversals
in the circular polarization and thus mark the
appearance/disappearance of one or both accreting regions over the
limb of the white dwarf, at which time the accreting field lines are
viewed perpendicularly from Earth. The variation in position angle is
poorly defined for most of the orbit, except at the phases of the
linearly polarized pulses.

\subsection{The system's geometry and dimensions}

We now use the eclipses and the system's parameters derived so far in order to
investigate the system's geometry and dimensions.

The mass and radius of the secondary star can be estimated by using
the mean empirical mass-period and radius-period relationships (Smith
\& Dhillon 1998) i.e.

$$ M2/M_{\sun} = (0.126 \pm 0.011)P - (0.11 \pm 0.04) = 0.1 (\pm 0.06) $$
and
$$ R2/R_{\sun} = (0.117 \pm 0.004)P - (0.041 \pm 0.018) = 0.16  (\pm 0.03)$$

Next, we use the fact that the duration of the total eclipse of the
primary is a function of the mass-ratio and the inclination only. This
cannot be expressed in an analytical form. However, Horne (1985)
presented a graphical representation of this relationship as a
function of eclipse width, which we now use to investigate the mass
ratio-inclination parameter space.

For stable mass transfer we require that $M_{2}/M_{1} < 1$ and,
therefore, using the above estimate of $M2 = 0.1M_{\sun}$, the mass of
the white dwarf primary must be $0.1 < M1 < 1.44 M_{\sun}$, which is
equivalent to $ 0.069 < M2/M1 < 1 $. Using this and the eclipse length
of $\Delta \phi = 0.029$ (derived from section 3.2) in conjunction
with figure 2 of Horne (1985), we find that the inclination must be in
the range $70^{\rm o} < i < 81.5^{\rm o}$. Propagating the errors for
the mass of the secondary increases the upper limit to $85^{\rm o}$.
We also estimate a range for the white dwarf radius (see Hamada \&
Salpeter 1961 or In-Saeng Suh \& Mathews 2000). These parameters are
tabulated in table~\ref{tab:sysparms} as a function of inclination.

\subsection{Modelling the cyclotron emission.}

\begin{figure}
\epsfxsize8.5cm
\epsffile{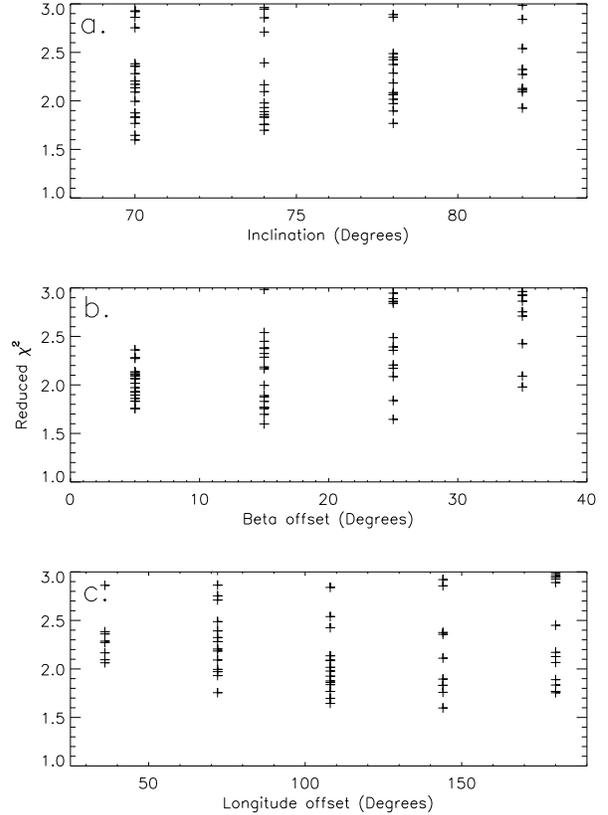} 
\caption{Final reduced $\chi^{2}$ from running Stokes imaging several
times, each assuming different values of the inclination and magnetic
dipole offset angles.  Results plotted as a function of a) inclination
b) the magnetic dipole offset ($\beta$) and c) the magnetic dipole
offset in longitude. }
\label{plotchis}
\end{figure}

\begin{figure*}
\epsfxsize=16.5cm
\epsffile{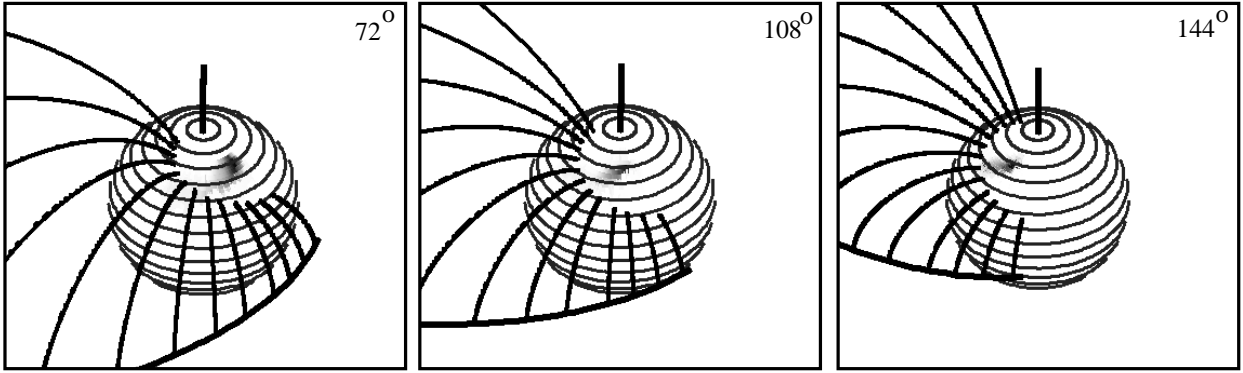} 
\caption{Predictions for the shape, size and location of the upper
accretion region (grey shaded area) assuming a binary inclination of
$70^{\rm o}$, magnetic dipole offset $\beta = 5^{\rm o}$ and magnetic
dipole longitude of $72^{\rm o}, 108^{\rm o}$ and $144^{\rm o}$ (from
left to right). The globes represent the surface of the white dwarf
with latitudes marked every 10 degrees. The upper magnetic pole is
indicated by a vertical straight line. The remaining solid curves
represent magnetic field lines. The white dwarf rotates in a
anti-clockwise direction as seen from above the upper spin axis. The
white dwarf is orientated such that the upper magnetic pole is most
parallel to the viewers line of sight. }
\label{inc70}
\end{figure*}

\begin{figure}
\epsfxsize=8.5cm
\epsffile{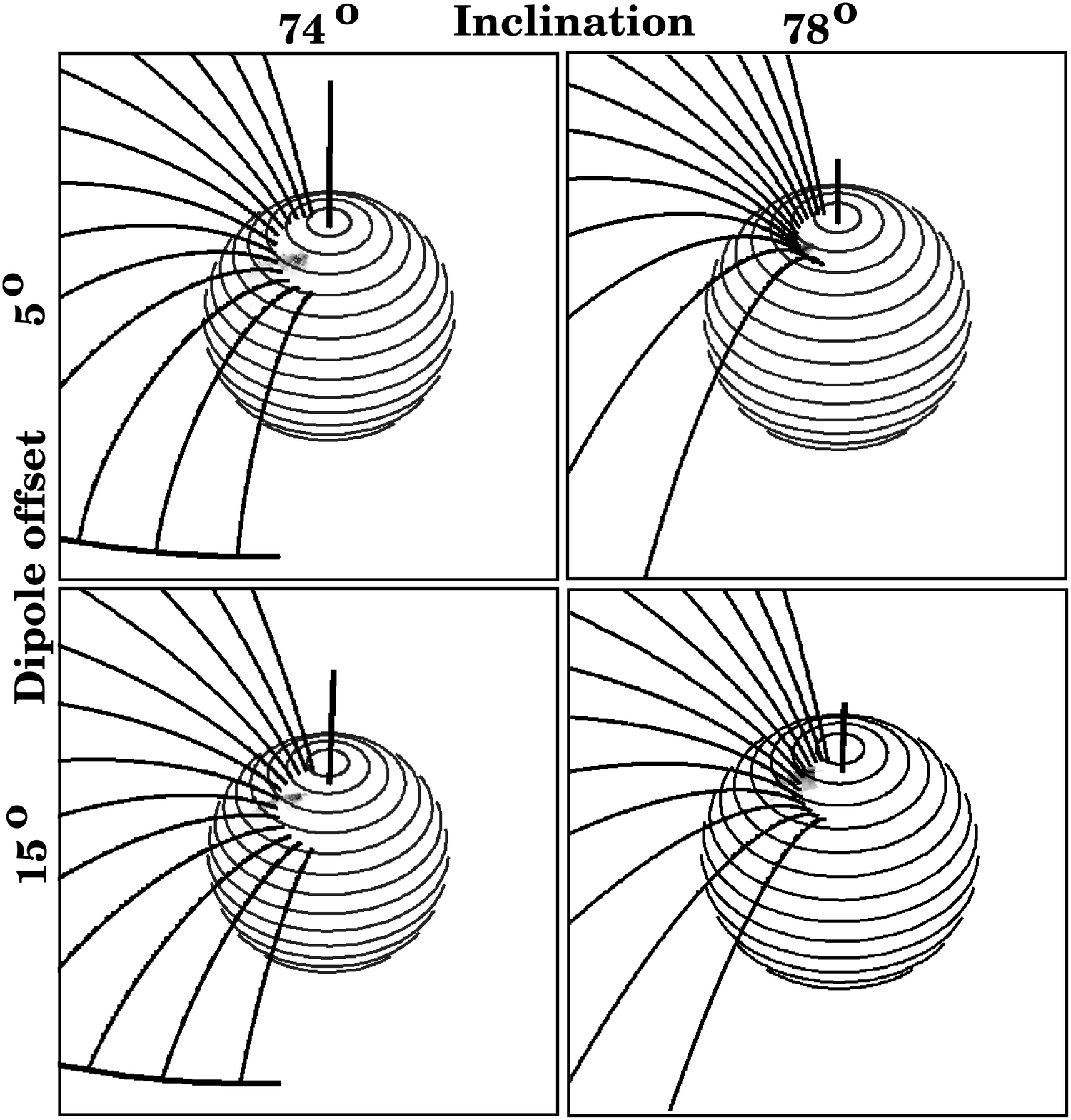}
\caption{As in Fig.~\ref{inc70} except, from left to right, binary
inclinations of $74^{\rm o}$ and $78^{\rm o}$ , and, from top to
bottom, magnetic dipole offset angles of $\beta = 5^{\rm o}$ and
$\beta = 15^{\rm o}$. A magnetic dipole longitude of $144^{\rm o}$ was
assumed in all cases.  }
\label{inc74,78}
\end{figure}

We next model the polarization data using the Stokes imaging
technique of Potter, Hakala \& Cropper (1998) and Potter et
al. (2004).  Stokes imaging allows the mapping of the accretion
region(s) (strictly speaking, the cyclotron emission regions) in
magnetic cataclysmic variables in terms of their location, shape and
size. Any particular solution depends on certain system's parameters
(inclination and the dipole offset angles in latitude and longitude),
so the technique was run several times in order to investigate the
solutions within the parameter space defined by the range in our
derived system's parameters.

The ranges in the binary inclination and the $\beta$ offset of the
magnetic pole were estimated in the previous sections and tabulated in
table~\ref{tab:sysparms}. We now estimate the longitude of the
magnetic pole (how far ahead it is from the line of centers of the two
stars) by assuming that the upper accretion region is located close to
the magnetic pole. Then, from Figs.~\ref{unfilt},\ref{OG570} \&
\ref{BG39}, it can be seen that the upper accretion region is eclipsed
by the secondary star towards the end of the bright phase. This is
$\sim 0.3$ of an orbit after the upper accretion region is most face
on to the viewer. This indicates that the upper accretion region, and
hence the upper magnetic pole, is ahead ($\sim 100^{\rm o}$ in
longitude) of the line of centers of the two stars.

Figs.~\ref{plotchis} a,b \& c show the final reduced $\chi^{2}$ from
running Stokes imaging several times, each assuming different values
of the inclination and magnetic dipole offset angles. The technique of
Potter, Hakala \& Cropper (1998) included an estimate of the
smoothness of the predicted emission region in the final fit. However,
we use the methodology outlined in Potter et al. (2004) in which
emission regions are smooth from the outset and therefore the final
reduced $\chi^{2}$ does not contain any extra terms. The three plots
refer to the same set of solutions, each plotted as a function of a
different parameter. The final reduced $\chi^{2}$ of the model fit to
the observations for a particular set of system parameters is
represented by a '+'. For each value of one of the parameters, there
is a range in $\chi^{2}$ corresponding to the fits for different
combinations of the other parameters.  Hence, each solution appears
once in every plot.

Formally, the best solution is obtained using an inclination of
70$^{\rm o}$ and dipole offset angles of 15$^{\rm o}$ and 144$^{\rm
o}$ for latitude and longitude, respectively. However, given the
shallowness of the minima in Fig.~\ref{plotchis}, we next investigate
the actual shape, size and location of the accretion regions on the
surface of the white dwarf predicted by Stokes imaging.

\subsubsection{The upper accretion region.}

We first concentrate on the solutions for the emission from the upper
accretion region only (the bright phase) because the polarized
emission from the upper accretion region is visible for a larger
fraction of the orbit than the lower accretion region. The visibility
of the upper accretion region is clearly demarcated by the positive
circular polarization and by the beginning and end points of the
bright phase in the photometry. Therefore, there is no ambiguity
associated with the visibility of the upper accretion region.

We now investigate the solutions with different values for the
longitudinal offset of the magnetic dipole. Fig.~\ref{inc70} shows
three solutions from Stokes imaging. From left to right, the magnetic
longitude was set to be $72^{\rm o}, 108^{\rm o}$ and $144^{\rm o}$.
Each solution assumed an inclination of $70^{\rm o}$ and a magnetic
dipole offset angle of $\beta = 5^{\rm o}$. The accretion region is
shown as the shaded grey area.

In each case we have overplotted a model ballistic trajectory computed
using a single particle under gravitational and rotational influences
only.  No additional drag terms (e.g. due to the magnetic field) are
included. The particle is allowed to follow a ballistic path from the
L1 point and is terminated 130 degrees in azimuth around the white
dwarf. At 10 degree intervals in azimuth (from 0 to 130 degrees)
around the white dwarf, dipole trajectories are calculated (the curved
diagonal lines) from the ballistic stream to the surface of the white
dwarf. The mass ratio and white dwarf mass were assumed to be as given
in table~\ref{tab:sysparms}. For a particular white dwarf mass and
inclination, the smaller white dwarf radius of the range defined in
table~\ref{tab:sysparms} was assumed.

We next compare the location of the footprints of the magnetic field
lines on the surface of the white dwarf with the prediction for the
location of the upper accretion region. This comparison serves to
check the validity of the predictions from Stokes imaging: the two
locations should be very similar.

The left and middle solutions of fig.~\ref{inc70} are rejected for two
reasons. Firstly, the location of the accretion region is found not to
be trailing the magnetic pole in orbital phase. The asymmetry in the
double-humped morphology of the bright phase in the photometry and, in
particular, in the positive circular polarisation suggests that the
accretion region trails the magnetic pole (see the discussion in
section 3.4 and Cropper 1998 for an example of where the accretion
region is thought to be ahead of the magnetic pole). Therefore, we
reject any of the solutions that predicted the upper accretion region
to be ahead of the upper magnetic pole. Secondly, because the
footprints of the magnetic field lines do not overlap with the
accretion region.

The right plot, however, shows the location of the accretion region to
be trailing the magnetic pole. The overlap between the magnetic
footprints and the accretion region is slightly better, but could not
be improved by increasing the longitude of the magnetic pole. In
addition, for angles larger than $180^{\rm o}$, the $\chi^{2}$ fit
becomes increasingly worse. We would like to point out that we have
used the smallest white dwarf radius possible as defined by the white
dwarf mass radius relationship (see Hamada \& Salpeter 1961 or
In-Saeng Suh \& Mathews 2000) and permitted by the estimate of the
system's parameters of table~\ref{tab:sysparms}. Increasing the white
dwarf radius decreases the overlap. 

Thus we can conclude that magnetic dipole longitudes smaller than
$\sim 144^{\rm o}$ are inconsistent with the modelling of the
polarimetry and incorrectly predict the accretion region to lead the
magnetic pole in orbital phase. This conclusion was reached by
assuming that accretion takes place from the unperturbed ballistic
stream threaded by unperturbed magnetic field lines, which is clearly
too simplistic. However, we argue that a more detailed model that
includes additional effects, such as magnetic drag, would give rise to
a stream that has smaller orbital velocities and would, most probably,
place it on a trajectory on the inside of the unperturbed ballistic
trajectory (see Heerlein, Horne \& Schwope 1999 for an investigation
into the effects of additional drag terms). Consequently, once
attached to the magnetic field lines, the footprints would be located
even further to the left of the accretion regions compared to those
shown in Fig.~\ref{inc70}.

Therefore, in what follows, the magnetic longitude is fixed at $\sim
144^{\rm o}$ and we investigate the solutions for different
inclinations and $\beta$ offsets.

We now examine the solutions as a function of inclination. For
inclinations less than 74 degrees, the magnetic footprints were found
to be always at lower latitudes than the accretion region predicted by
Stokes imaging, as already demonstrated in Fig.~\ref{inc70}. For
inclinations of $\sim 74^{\rm o}$ the overlap was marginal, and it's
best for inclinations of $\sim 78^{\rm o}$ (Fig.~\ref{inc74,78}). This
is towards the middle of the range in inclination estimated from using
the eclipse length in combination with figure 2 of Horne (1985) (see
section 3.5 above). Inclinations of $\sim 78^{\rm o}$ are also more in
agreement with the implied mass ratios and primary masses
(table~\ref{tab:sysparms}) expected for CVs with similar orbital
periods (Smith \& Dhillon 1998).  We have assumed that the difference
in primary masses between magnetic and nonmagnetic white dwarfs to be
insignificant assuming the magnetic field strength used here (see
In-Saeng Suh \& Mathews 2000).

Fig.~\ref{inc74,78} also shows the solutions for different values for
the $\beta$ offset of the magnetic dipole. Good overlap between the
magnetic footprints and the predicted location of the upper accretion
region from Stokes imaging is found for $\beta$ angles of 5 and
15$^{\rm o}$. The reduced $\chi^{2}$ fit, as a function of $\beta$
offset angle (Fig.~\ref{plotchis} b), also predicts best values within
this range.


\subsubsection{The lower accretion region.}

\begin{figure}
\epsfxsize=8.5cm
\epsffile{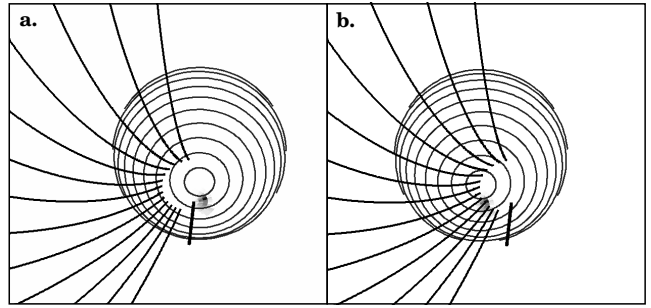} 
\caption{a. Predictions for the shape, size and location of the lower
accretion region viewed from above the lower spin axis of the white
dwarf. b. As a) except the lower magnetic pole has been moved in order
to `force' the magnetic field lines to overlap with the lower accretion
region (see text) }
\label{lowpole}
\end{figure}

As described earlier, it is more difficult to be certain of the phases
at which this accretion region is in view. However, we can assume that
the lower accretion region is in view for a phase range at least
defined by the appearance of the negative circular polarization.

The same values for the inclination, mass-ratio and white dwarf mass
that gave the best results for the upper accretion region were
used. However, we found that by using the same dipole geometry for the
upper accretion region, the lower footprints of the magnetic
fieldlines, from the same ballistic stream, did not coincide with the
lower accretion region predicted by Stokes imaging (Fig.~\ref{lowpole}
a). Not surprisingly, this can be accounted for by assuming that the
magnetic field of the white dwarf is not simply a tilted
dipole. Fig.~\ref{lowpole} b. shows that by offsetting the lower
magnetic pole in longitude and latitude by $\sim 30^{\rm o}$ and $\sim
10^{\rm o}$ respectively, whilst keeping the upper magnetic pole
fixed, the magnetic footprints can be `forced' to coincide with the
lower accretion region. However, it should be noted that, as mentioned
above, the location of the lower accretion region may be incorrect due
to the possible lack of phase coverage given by the negative
polarisation.

If the magnetic field geometry were different from that of a simple
dipole, it would not be possible to obtain a unique magnetic field
geometry by modelling both accretion regions simultaneously.  However,
the polarized emission from {\it one} accretion region located close
to a magnetic pole is only weakly dependent of the magnetic field
topology of the white dwarf as a whole. This is because the
inclination of the magnetic field lines close to a magnetic pole are
very similar for most magnetic field topologies. Therefore, the
conclusions drawn in section 3.6.1 for the upper accretion region
would not be expected to require significant modification.



\subsubsection{The model light curves.}

The model light curves (for both accretion regions) for an inclination
of 74$^{\rm o}$, magnetic dipole angles of 144$^{\rm o}$ and 5$^{\rm
o}$ are shown overplotted on the polarimetric observations in
Fig.~\ref{unfilt}. Note that the aim of Stokes imaging is to fit the
general morphological variations of the observations, but not to fit
the small scale details that either cannot be distinguished from
noise, or that go beyond the assumptions of the cyclotron
model. Hence, as can be seen from Fig.~\ref{unfilt}, the model has fit
well the general variations of the light curves, namely the
double-humped bright phase (intensity and circular polarisation), the
faint phase and the positions of the linear pulses. The model fit is
worse during phases $\sim 0.7 - 0.1$, where too much circular
polarization is predicted. This may arise because the model does not
take into account any additional absorption by the accretion curtain
that intersects our line of sight during these phases.

\section{Summary and Conclusions}

We have found polarisation in the new eclipsing polar CTCV~J1928-5001,
thus identifying it as a polar with an orbital period of $\sim$ 101
minutes. Our polarimetric observations show that there are two
accretion regions and the white dwarf has a magnetic field strength of
$\sim$ 20 MG.

From the eclipses we estimate some of the system's parameters, which we
further refine by modelling the polarimetric observations and making
comparisons with calculations of single particle ballistic and
magnetic trajectories.

We find the inclination to be 77$^{\rm o}\pm 2^{\rm o}$, the offset of
the upper magnetic pole from the spin axis to be between $\sim
5-15^{\rm o}$ and to be no less than $\sim 144^{\rm o}$ ahead of the
line of centers of the two stars. The single particle ballistic models
agree well with the results from Stokes imaging for the location of
the upper accretion region within this parameter range. Agreement is
not found for the location of the lower cyclotron accretion
region. This could either be due to the lack of polarization from the
lower accretion region, due to the dominance of emission from the upper
accretion region, or to the assumption of a centered dipole being
incorrect. Both reasons could also apply simultaneously. The presence
of two sets of cyclotron harmonics from spectroscopic observations
(e.g. Schwope et al. 1995) could help to determine the magnetic
topology (e.g. an offset dipole) of the white dwarf.

\section{Acknowledgments}

Thanks go to E. Romero-Colmenero for valuable discussions. We would
also like to thank the Referee for comments which significantly
improved the paper.

\end{document}